\documentclass{aa}  
\usepackage{graphicx}
\usepackage[varg]{txfonts}

\usepackage{natbib}
\bibpunct{(}{)}{;}{a}{}{,}
\begin{document}
\title{Young stars and reflection nebulae near the lower ``edge'' of the Galactic molecular disc\thanks{Based on observations collected at the ESO 8.2-m VLT-UT1 Antu telescope (programme 66.C-0015A).}
}


   \author{Pedro M. Palmeirim
		\and
          Jo\~ao L. Yun
          }


   \institute{Universidade de Lisboa - Faculdade de Ci\^encias \\
Centro de Astronomia e Astrof\'{\i}sica da Universidade de Lisboa, \\
Observat\'orio Astron\'omico de Lisboa, \\
Tapada da Ajuda, 1349-018 Lisboa, Portugal\\
             \email{ppalm@oal.ul.pt, yun@oal.ul.pt}
             }

   \date{Received September 1, 2009; accepted November 16, 2009}


 
  \abstract
   {}
   {We investigate the star formation occurring in a region well below the Galactic plane towards the optical reflection nebula ESO~368-8 (IRAS~07383$-$3325). We confirm the presence of a small young stellar cluster  (or aggregate of tens of YSOs) identified earlier, embedded in a molecular cloud located near the lower ``edge'' of the Galactic disc, and characterise the young stellar population. We report the discovery of a near-infrared nebula, and present a CO map revealing a new dense, dynamic cloud core.
}
   {We used near-infrared $JHK_S$ images obtained with VLT/ISAAC, millimetre CO spectra obtained with the SEST telescope, and optical $V$-band images from the YALO telescope.}
   {
This star formation region displays an optical reflection nebula (ESO~368-8) and a near-infrared nebula located about 46$''$ (1.1 pc) from each other. The two nebulae are likely to be coeval and to represent two manifestations of the same single star formation episode with about 1~Myr age. The near-IR nebula reveals an embedded, optically and near-IR invisible source whose light scatters 
off a cavity carved by previous stellar jets or molecular outflows and into our line-of-sight.
The molecular cloud is fully covered by our CO($J$=1$-$0) maps and, traced by this line, extends over a region of $\sim 7.8 \times 7.8$~pc$^2$, exhibiting an angular size $\sim 5.4' \times 5.4'$ and shape (close to circular) similar to spherical (or slightly cometary) globules.
Towards the direction of the near-IR nebula, the molecular cloud contains a dense core where the molecular gas exhibits large line widths indicative of a very dynamical state, with stirred gas and supersonic motions.
Our estimates of the mass of the molecular gas in this region range from 600 to 1600~$M_{\odot}$.
The extinction $A_V$ towards the positions of the optical reflection nebula and of the near-IR nebula was found to be $A_V \simeq 3-4$~mag and  $A_V \simeq 12-15$~mag, respectively.
}
   {}


\titlerunning{Young stars and reflection nebulae...}
\authorrunning{Palmeirim \& Yun}

   \maketitle
%

\section{Introduction}

Since early years, optical reflection nebulae have attracted much attention by their extended morphology and clearly non-point-like appearance \citep[e.g.][]{hubble22, barnard13, barnard27, vandenbergh66}. In the vicinity of reflection nebulae, different types of objects have been found, many associated with young stars and the process of star formation (e.g. emission-line stars, Herbig-Haro objects, jets and outflows, etc.). For a considerable period, these nebulae were among the best tracers of recent star formation ocurring in a region of the sky. 

Subsequently, many infrared star clusters and stellar groups have been found during surveys towards directions of optical and radio HII regions. \citet{testi98} found sources around Herbig AeBe stars, part of them related to reflection nebulae.
\citet{soares02,soares03} studied four low-mass star clusters in the CMaR1 molecular cloud, three of them related to optical reflection nebulae. \citet{dutra03} and \citet{bica03} investigated more than 3000 optical and radio nebulae with 2MASS \citep{skrutskie97} and found 346 new embedded star clusters, stellar groups, and candidates. 

Star formation sites are commonly associated with molecular material belonging to the parental cloud out of which the young stars form. 
As the young stars evolve, they feed from the available gas and dissipate it. 
Embedded clusters emerge from molecular clouds owing to the disruptive action of newly born stars \citep{matzner00}. Jets and outflows carve cavities into the surrounding molecular material. Scattering off the walls of these cavities allows photons from embedded young stellar objects (YSO) to escape and be detected as near-IR nebulosity, denouncing the presence of one or more YSOs \citep[e.g.][]{yun93,whitney93,whitney03}. 

Furthermore, since most gas and dust that exist in the Galaxy are confined to the Galactic disc, and are most abundant toward the inner Galaxy \citep{clemens88}, it is of particular interest to search for and characterise the star formation ocurring near the ``edges'' of the molecular disc, either along the Galacic disc in the far outer Galaxy or at large vertical distances from the Galactic disc. 

As part of our study of young embedded clusters in the third Galactic quadrant \citep[e.g.][]{yun07,yun09}, we report here an investigation of a relatively unstudied star formation region (SFR), seen towards the IRAS source IRAS~07383-3325. The position of the IRAS source coincides with that of an optical reflection nebula. The nebula is evident in the Digitized Sky Survey images (especially in that corresponding to the blue O-print) and was discovered by \citet{lauberts82}, who named it ``ESO~368-8'' and at that time classified it as an emission nebula. It is also listed in the catalogues of HII regions and reflection nebulae of \citet{brand86}  and of  \citet{magakian03} (and identified under the names ``BRAN~61'', ``Magakian 287'', and ``GN~07.38.4''). 

Optical examination of the Palomar and ESO plates \citep{brand86} did not find any evidence of obscuration towards ESO~368-8. Thus, the molecular nature of this SFR was first revealed in the CO survey of \citet{wouterloot89} in their survey of IRAS sources beyond the solar circle, where IRAS~07383-3325 is listed as a good detection (no. 1039). We refer to this molecular cloud as WB89-1039, first mapped and probed by us using millimetre spectra, and presented here in Sect.~\ref{distance}.

A preliminary search for the young stellar content of an SFR can be performed using 2MASS images in the Ks-band. This has been done by \citet{dutra03}, who indentified a candidate ``open cluster (IROC)'' (DBSB~15) located about $22''$ east of the IRAS source with major and minor axes of $1.9'$ and $1.7'$. 
Here, using a combination of deep near-infrared ($JHK_S$-bands) and optical ($V$-band) images, we characterise the stellar content of this SFR towards the reflection nebula ESO~368-8/IRAS~07383-3325 and molecular cloud WB89~1039. We confirm the young stellar nature of the candidate open cluster of \citet{dutra03}. We found a group of stars, spatially coincident with DBSB~15 and extending about 1$^{\prime}$ to the east of the IRAS source, involved in infrared nebulosity. We thus report the discovery of a near-infrared reflection nebula seen in the near-IR images. 

The Galactic coordinates of this SFR are $l=248.01^{\circ}$, $b=-5.46^{\circ}$ which, together with its estimated distance of 5 kpc (see Sect.~\ref{distance}), places this region well below the Galactic plane in the region of the Galactic warp \citep{momany06,may97}. Thus, this study adds to what has been our relatively poor knowledge of star formation occurring in ``frontier'' environments, in this case, the lower ``edge'' of the Galactic disc.

Section~2 describes the observations and data reduction. In Sect.~3, we present and analyse the results. Section~4 discusses the star formation scenario in this region. A summary is given in Sect.~5.


\section{Observations and data reduction}

\subsection{Near-infrared observations}

Near-infrared ($J$, $H$ and $K_S$) images were obtained on 2000 November 11 using the ESO Antu (VLT Unit 1) telescope equipped with the short-wavelength arm (Hawaii Rockwell) of the ISAAC instrument. The ISAAC camera \citep{moorwood98} contains a 1024 $\times$ 1024 pixel near-infrared array and was used with a plate scale of 0.147 arcsec/pixel resulting in a field of view of 2.5~$\times$~2.5 arcmin$^2$ on the sky.
For each filter, 6 dithered sky positions were observed. A series of 15 images with individual on-source integration time of 4 seconds was taken in the $J$-band. Similarly, series of 20 images, each of 3 second integration time, were obtained in the $H$ and the $K_S$ bands. 

The images were reduced using a set of own IRAF scripts to correct for 
bad pixels, subtract the sky background, and flatfield the images. Dome flats were used to correct for the pixel-to-pixel variations of the response. The selected images were 
then aligned, shifted, trimmed, and co-added to produce a final mosaic image for each band $JHK_S$. 
Correction for bad pixels was made while constructing the final mosaics that cover about $3.8 \times 3.1$ arcmin$^2$ on the sky.

Point sources were extracted using {\tt daofind} with a detection threshold of 
5$\sigma$. The images were inspected to look for false detections that had been included by {\tt daofind} in the list of detected sources. These sources were eliminated from the source list and a few additional sources added in by hand. Aperture 
photometry was made with a small aperture (radius = 3 pix, which is about
the measured FWHM of the point spread function) and aperture corrections, 
found from bright and isolated stars in each image, were used to correct 
for the flux lost in the wings of the PSF. The error in determining 
the aperture correction was $<$ 0.05 mag in all cases. 
A total of 900, 860, and 969 sources were found to have fluxes in $J$, 
$H$, and $K_S$, respectively, and errors $<$ 0.15 mag.

The $JHK_S$ zeropoints were determined using faint infrared standard stars  \citep[S427-D and S121-E of the LCO/Palomar NICMOS Table of Photometric Standards; see][]{persson98} and checked with those obtained using 2MASS stars brighter than $K_S$ = 13.5 mag from the 2MASS All-Sky Release Point Source Catalogue \citep{skrutskie06,cutri03}.
Photometry errors range from about 0.08 mag for the bright stars to 0.15 mag for the fainter ones. From the histograms of the $JHK_S$ magnitudes, we estimate the completeness limit of the observations to be roughly 20.0 magnitudes in the $J$-band, and 18.5 magnitudes in the $H$ and $K_S$ bands.

\subsection{CO mm line observations}
Millimetre single-dish observations of IRAS~07383-3325 were carried out at the 15~m 
SEST telescope (ESO, La Silla, Chile) during three observational campaigns 
in 2001~May, 2002~April, and 2002~December. 
Starting from the (0,0) reference position corresponding to the celestial coordinates of the IRAS source, 
maps were obtained in the rotational lines of $^{12}$CO(1$-$0) (115.271~GHz), 
$^{13}$CO(1$-$0) (110.201~GHz), and $^{13}$CO(2$-$1) (220.399~GHz), C$^{18}$O(1$-$0) (109.782~GHz), and C$^{18}$O(2$-$1) (219.560~GHz). 
The SEST half-power beam width (HPBW) was $46\arcsec$ at 110~GHz 
and $22\arcsec$ at 230~GHz, so that, since the adopted grid spacing was 
$46\arcsec$, the $^{13}$CO(2$-$1) and the C$^{18}$O(2$-$1) line emissions were strongly undersampled. 
The $^{12}$CO(1$-$0) and $^{13}$CO(2$-$1) maps are composed of $8 \times 9$ pointings, 
while the $^{13}$CO(1$-$0), C$^{18}$O(1$-$0), and C$^{18}$O(2$-$1) maps are smaller ($5\times 5$, $6\times 5$, and $3\times 3$ pointings, respectively).

The adopted integration times were between $t_\mathrm{int}=60$~s and $t_\mathrm{int}=240$~s for each pointing. 
A high-resolution 2000 channel acousto-optical spectrometer was used as a
back end, with a total bandwidth of 86~MHz and a channel width of 43~kHz, and 
was split into two halves to measure both the 115 and 230~GHz receivers 
simultaneously 
At these frequencies, the channel 
width corresponds to approximately 0.11~km~s$^{-1}$ and 0.055~km~s$^{-1}$. 
The spectra were taken in frequency switching mode, 
recommended to save observational time when mapping extended sources.
The antenna temperature was calibrated with the standard chopper wheel method.
Pointing was checked regularly towards known circumstellar
SiO masers, and pointing accuracy was estimated to be better than
$5\arcsec$.

The data reduction pipeline was composed of the following steps: 
\textit{i)}  folding the frequency-switched spectrum; 
\textit{ii)} fitting the baseline by a polynomial and subtracting it; 
\textit{iii)} co-adding repeated spectra obtained at the same sky position, weighing with the integration time and the inverse of the system temperature; 
\textit{iv)} obtaining the main-beam temperature $T_{\mathrm{MB}}$ by dividing the antenna temperature $T_{\mathrm{A}}$ by the $\eta_{\mathrm{MB}}$ factor, equal to 0.7 for the lines at 110-115~GHz and 0.5 for the other lines (at 220~GHz);

The spectrum baseline RMS noise (in $T_{\mathrm{MB}}$), averaged over 
all map positions, has been found to be 0.13~K for $^{12}$CO(1$-$0), 0.09~K for
$^{13}$CO(1$-$0), 0.15~K for $^{13}$CO(2$-$1), 0.04~K for C$^{18}$O(1$-$0), and 0.11~K for C$^{18}$O(2$-$1).

\subsection{Optical $V$-band observations}
Our optical $V$-band observations were conducted during
1999 December using the 1m YALO\footnote{YALO is the {\bf Y}ale-{\bf A}URA-{\bf 
L}isbon-{\bf O}hio consortium \citep{bailyn99}} telescope, at the Cerro Tololo 
Inter-American Observatory in Chile, equipped with the ANDICAM camera. 
The ANDICAM CCD camera contained 2048 $\times$ 2048 pixels, but read-out problems meant that only about 1400 $\times$ 1024 pixels were usable.
On a plate scale of 0.3 arcsec/pixel, it provided a field of view of 
7.0~$\times$~5.1 arcmin$^2$ on the sky.
Standard procedures for CCD image reduction were applied.
Source extraction and aperture photometry were performed using IRAF packages.
Transformation to the standard system was made using a fit to stars common to our images and to the US Naval Observatory Catalog. The results were checked using the spectrophotometric data of \citet{fitzgerald76}. Their stars no. 27 and 28 fall in our $V$-band image and their magnitudes are compatible (within 0.1 mag) with the results of our photometry.

\section{Results and discussion}

\subsection{Optical nebula and stars}

\begin{figure}
   \centering
   \includegraphics[width=8cm]{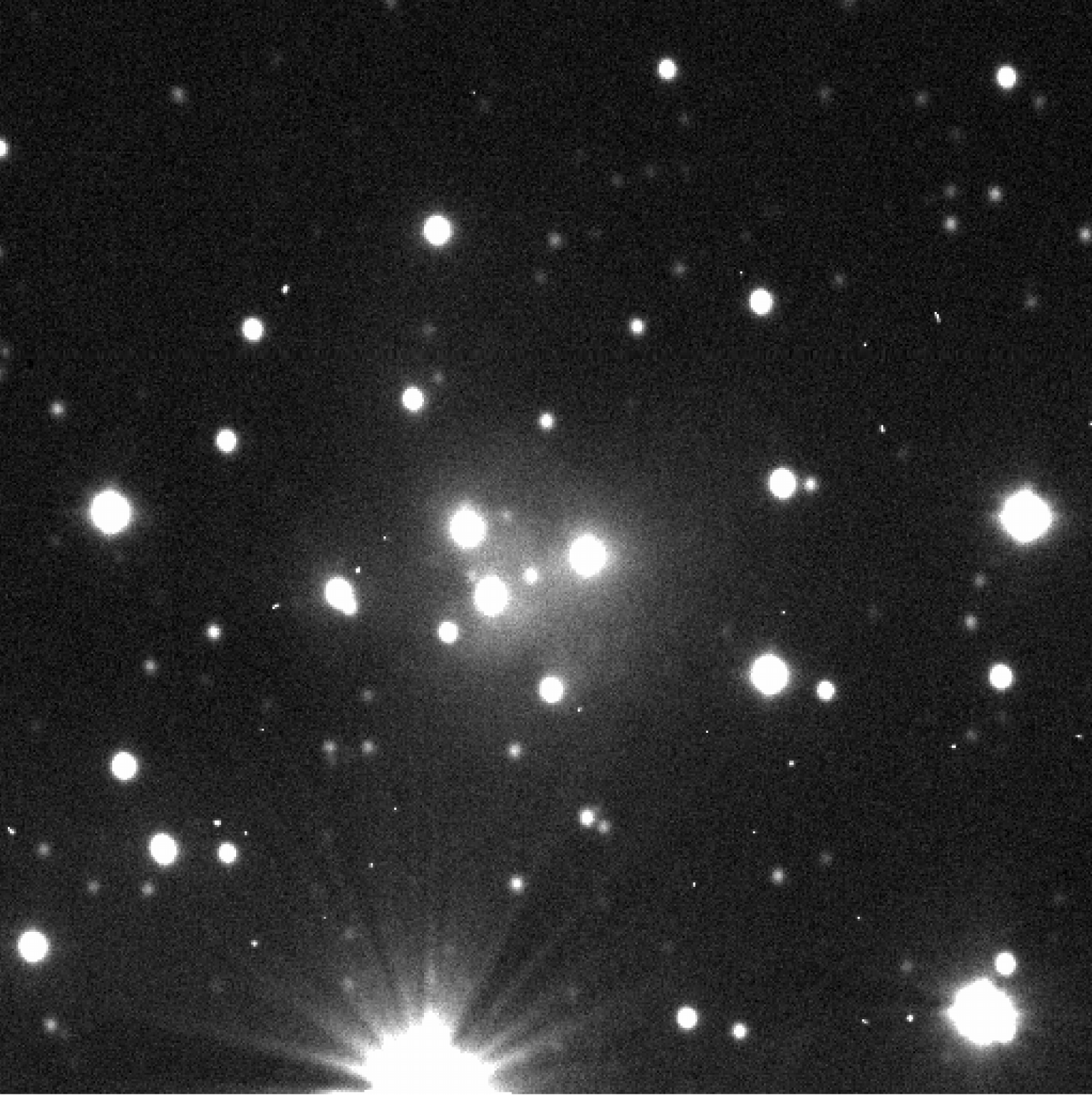}
   \caption{
Central region of the $V$-band image towards ESO~368-8/IRAS~07383-3325 (corresponding approximately to the region of the cluster
DBSB~15, proposed by \citet{dutra03}). Notice the central group of stars involved in optical nebulosity. 
This region covers about $2.56\times 2.56$ arcmin$^2$ and is centred on the IRAS source. North is up and east to the left.
	} 
	\label{i07383v4}%
    \end{figure}

\begin{figure} 
   \centering
   \includegraphics[width=8cm]{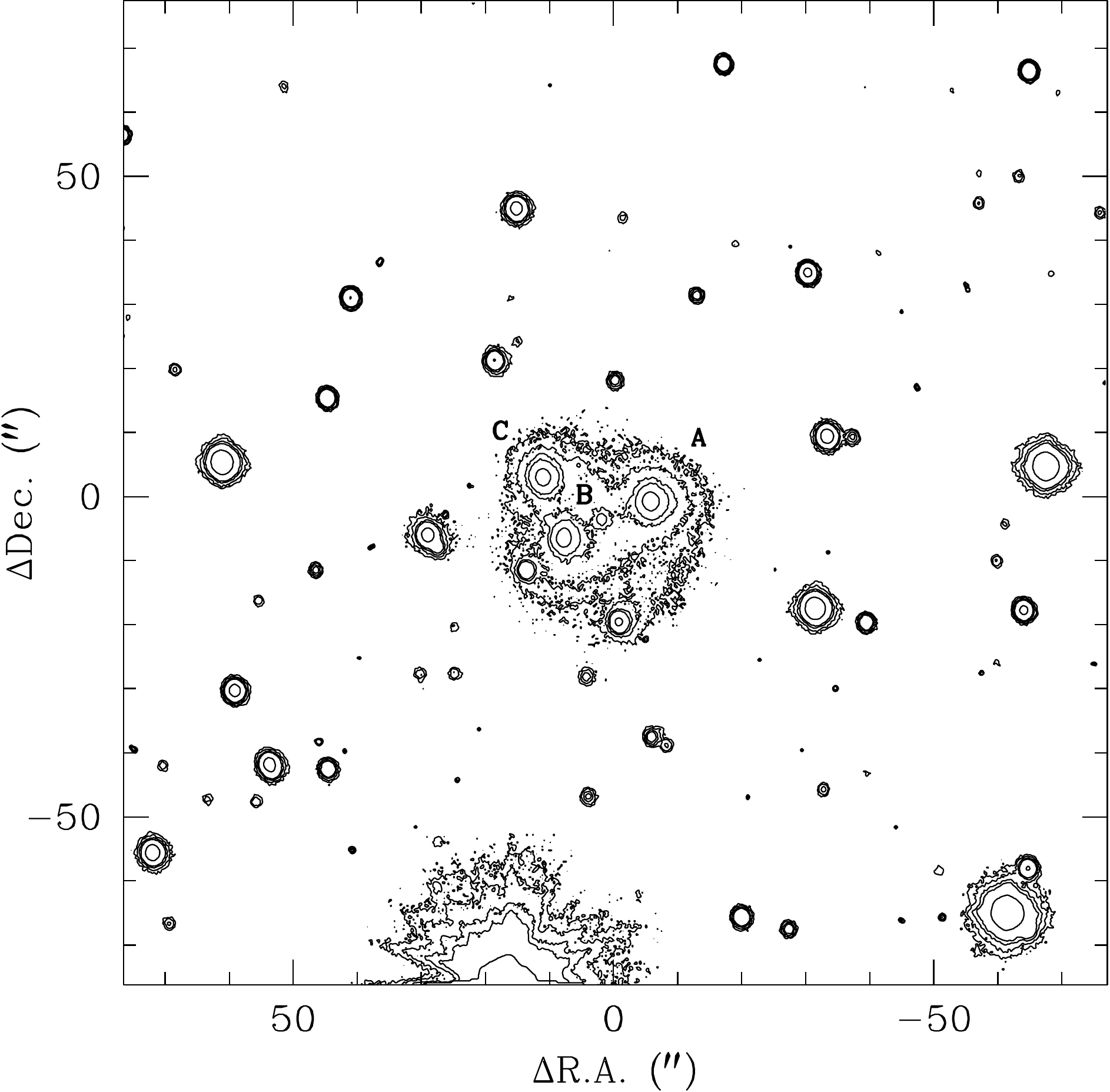}
   \caption{
Isophotes of the $V$-band image towards ESO~368-8/IRAS~07383-3325, with labels (A, B, and C) referring to the 3 brightest optical nebula stars (see text). The contour levels are 8, 12, 20, 30, and 200$\sigma$ above the background sky value. The IRAS source is at the (0,0) position.
	} 
	\label{v4contour}%
    \end{figure}

Figure~\ref{i07383v4} shows the central region of the $V$-band image towards ESO~368-8/IRAS~07383-3325. 
At optical wavelengths, this region of the sky does not look particularly spectacular which justifies the relatively little attention that has been devoted to it. 
At the centre of the image, a group of a few stars is seen towards optical nebulosity.
Basically, this optical nebula appears to clearly involve 3 stars with similar magnitudes and possibly one or two fainter stars. 
This can be more clearly seen in Fig.~\ref{v4contour} where the 3 stars are labelled A, B, and C.
The position of the IRAS source lies among this group of optical stars (but does not coincide with any) and well within the optical nebula. We name these stars here as ``the optical nebula stars''. On the other hand, the IRAS ellipse error (not drawn in the figure) encompasses more than one star. Given the large IRAS beams, it is likely that all of these optical nebula stars contribute to the IRAS fluxes.

We note that, in our images, the optical nebula ESO~368-8 extends for about $1'$ diametre (extension measured down to a level of 8$\sigma$ above the background sky brightness; down to 3$\sigma$, the size increases just slightly, to $70''$), much smaller than the quoted size of $4'$ by \citet{brand86}.

%

Table~1 gives our photometry of the three brightest optical nebula stars in the optical $V$-band and in the near-infrared $JHK_S$-bands.

\begin{table*}[ht]
\begin{minipage}[t]{\columnwidth} 
\caption{Photometry of the optical nebula stars }
\label{optical}
\renewcommand{\footnoterule}{}  
\begin{tabular}{lccccccc}
\hline\hline
ID & Multiplicity & R.A. & Dec & m$_V$ & m$_{K\!_S}$ & $(H-K_S)$ & $(J-K_S)$ \\
 & & (2000) & (2000) & & & &  \\
\hline
A  & $\ldots$ & 07 40 14.7  & -$\,$33 32  34   & 14.90  & 12.46 & 0.24 & 0.65 \\
B\footnote{This star is double, as revealed by our near-IR images. The $V$-band magnitude quoted here corresponds to the combined flux of both stars, unresolved in the optical image} & 1 & 07 40 15.8  & -$\,$33 32  39   & 15.27  & 13.04 & 0.29 & 0.95 \\
 & 2 & 07 40 15.8  & -$\,$33 32  40   & 15.27  & 12.36 & 0.26 & 0.76 \\
C & $\ldots$ & 07 40 16.1  & -$\,$33 32  31      & 15.20  & 13.66 & 0.16 & 0.45 \\
\hline
\end{tabular}
\end{minipage}
\end{table*}


\subsection{Molecular gas and distance} \label{distance}

\begin{figure}
   \centering
   \includegraphics[width=8cm]{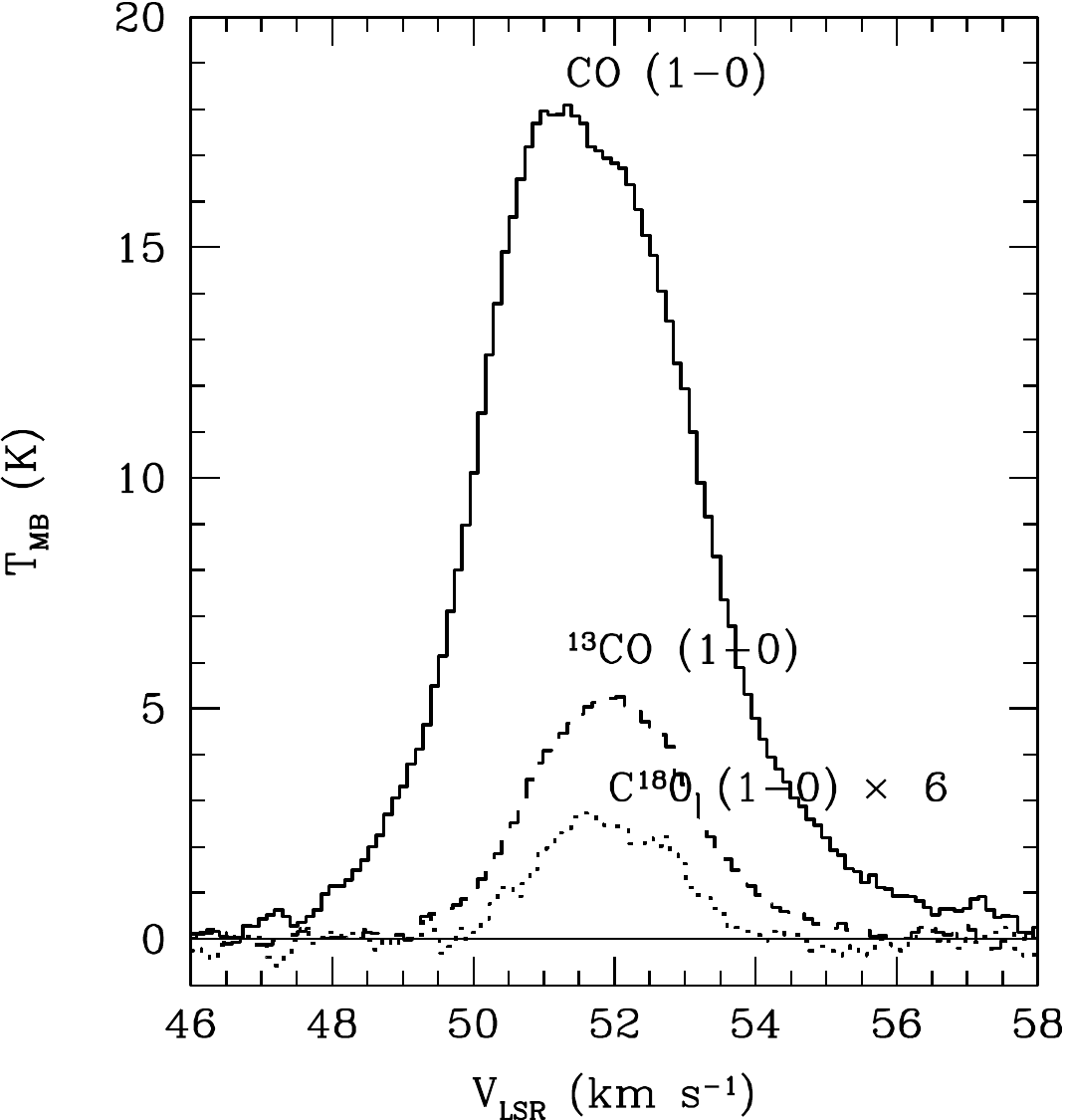}
   \caption{
J=1-0 spectra of CO isotopes towards the position of the map where the CO emission is strongest (about 46$''$ east of IRAS~07383-3325). 
The $^{12}$CO(1$-$0) line peaks at an LSR velocity of about 51.3~km~s$^{-1}$. 
The C$^{18}$O(1$-$0) line has been multiplied by a factor 6.
	} 
	\label{spectra}%
    \end{figure}

CO emission has been detected toward this region confirming the presence of a molecular cloud.
In Fig.~\ref{spectra}, we present the CO (J=1-0), $^{13}$CO (J=1-0), and C$^{18}$O (J=1-0) spectra obtained towards the position of the map where the CO emission is strongest: the $(+46,0)$ position of the map, or about 46$''$ east of IRAS~07383-3325. CO lines are detected with LSR velocities in the range of $V_{\mathrm{LSR}} \simeq 47 - 57$~km~s$^{-1}$.
The $^{12}$CO(1$-$0) spectrum peaks at $V_{\mathrm{LSR}} = 51.3 $ km~s$^{-1}$, while those of $^{13}$CO, and C$^{18}$O lines peak at 52.1 and 51.6~km~s$^{-1}$, respectively; these values represent the velocities of the gas component in this molecular cloud.

%

The kinematic heliocentric distance of this cloud can be derived from the peak radial velocities of the spectra. This can be done applying the circular rotation model by \citet{brand93}.
Using $V_{\mathrm{LSR}}=51.3$~km$^{-1}$, a distance $d\simeq 5.0$~kpc is obtained.
Given the Galactic longitude, this implies a Galactocentric distance of 11.4~kpc to the cloud, in the outer Galaxy. We estimate a distance uncertainty of up to 20\% due to uncertainties in the rotation curve and possible streaming motions. 

Given the Galactic latitude, the numbers above imply a vertical distance to the Galactic plane (defined by $b=0^{\circ}$) of about 500~pc, well below it and within the region of the Galactic warp \citep{may97}.
The spatial location of this molecular cloud, both at a relatively large heliocentric distance in the outer Galaxy and far below the Galactic plane, suggests that there are not many background stars (behind the cloud), as discussed in section \ref{reddening}.

\subsubsection{Molecular gas distribution: a new dense cloud core}

\begin{figure}
\centering
 \includegraphics[width=8cm]{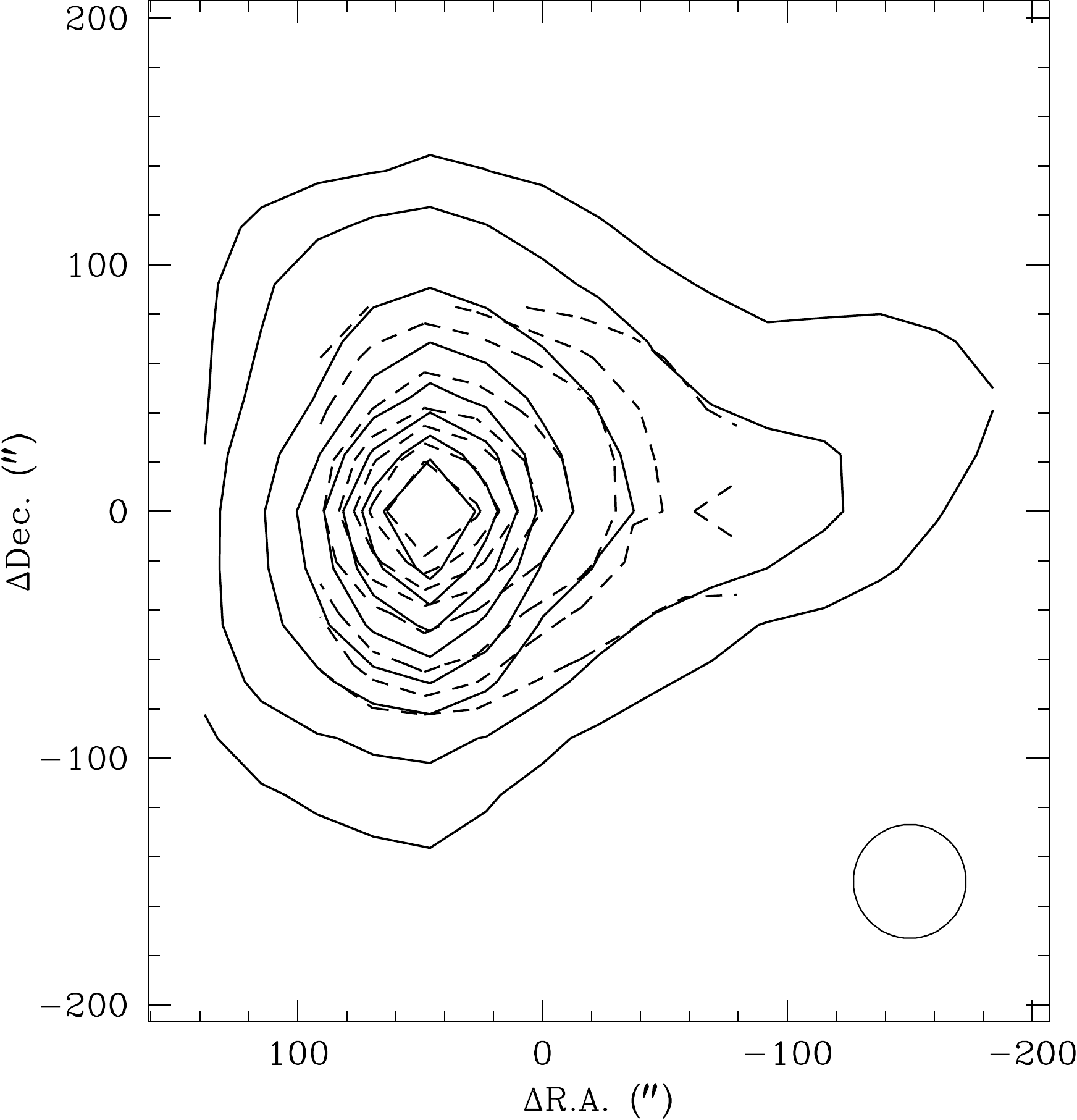}
\caption{
Smoothed contours of $^{12}$CO(1$-$0) (solid line) and of $^{13}$CO(1$-$0)  (dashed line) integrated intensities, in the velocity range between 47 and 57 km~s$^{-1}$, observed in the WB89-1039 and IRAS~07383-3325 region, containing a dense core (PY10) about $46''$ east of the IRAS (0,0) position. The two $^{12}$CO(1$-$0) lowest contour levels are at 1~K~km~s$^{-1} (5\sigma)$ and 3~K~km~s$^{-1}$. Subsequent contours are in steps of 5~K~km s$^{-1}$. 
The three $^{13}$CO(1$-$0) lowest contour levels are at 0.3~K~km~s$^{-1}$ (2.5$\sigma$) and 0.6~K~km~s$^{-1}$ and 1~K~km~s$^{-1}$. Subsequent contours  are in steps of 1~K~km~s$^{-1}$.
The SEST beam at 115 GHz is displayed in the lower right corner.
The IRAS source is at the (0,0) position.
}
\label{maps}
\end{figure}

The contour maps of the $^{12}$CO(1$-$0) and $^{13}$CO(1$-$0) integrated intensities $I=\int T_{\mathrm{MB}}~dv$ are shown in Fig.~\ref{maps}.
The CO emission is resolved and confined to the mapped region. 
Our CO maps presented here represent the totality of the WB89-1039 cloud, observed with unprecedented resolution.  
Only lines characterized by a signal-to-noise ratio $>3$ were 
considered as genuine emission. 
The cloud appears slightly elongated approximately in the direction east-west. In the eastern part, there is a peak of strong emission in a core-like or nucleus region; in the western part of the cloud, the emission is narrower and more diffuse in a tail-like fashion. Thus, we identify a new dense cloud core that we designate by ``PY10 dense core''.

In all maps, the CO emission peaks at the $(+46,0)$ (about 46$''$ east of IRAS~07383-3325), and decreases steeply in all directions. The shape of the strong CO emission is thus approximately circular and centred on the (+46,0) position. We name this position the ``WB89-1039 cloud centre'', or the ``PY10 core centre''. This cloud position is where the gas is warmest and has the highest column density.
The maps of C$^{18}$O(1$-$0) and C$^{18}$O(2$-$1) are not shown because detection of these lines is practically unresolved, restricted to one map position only, the PY10 centre position.

\begin{figure}
\centering
 \includegraphics[width=8cm]{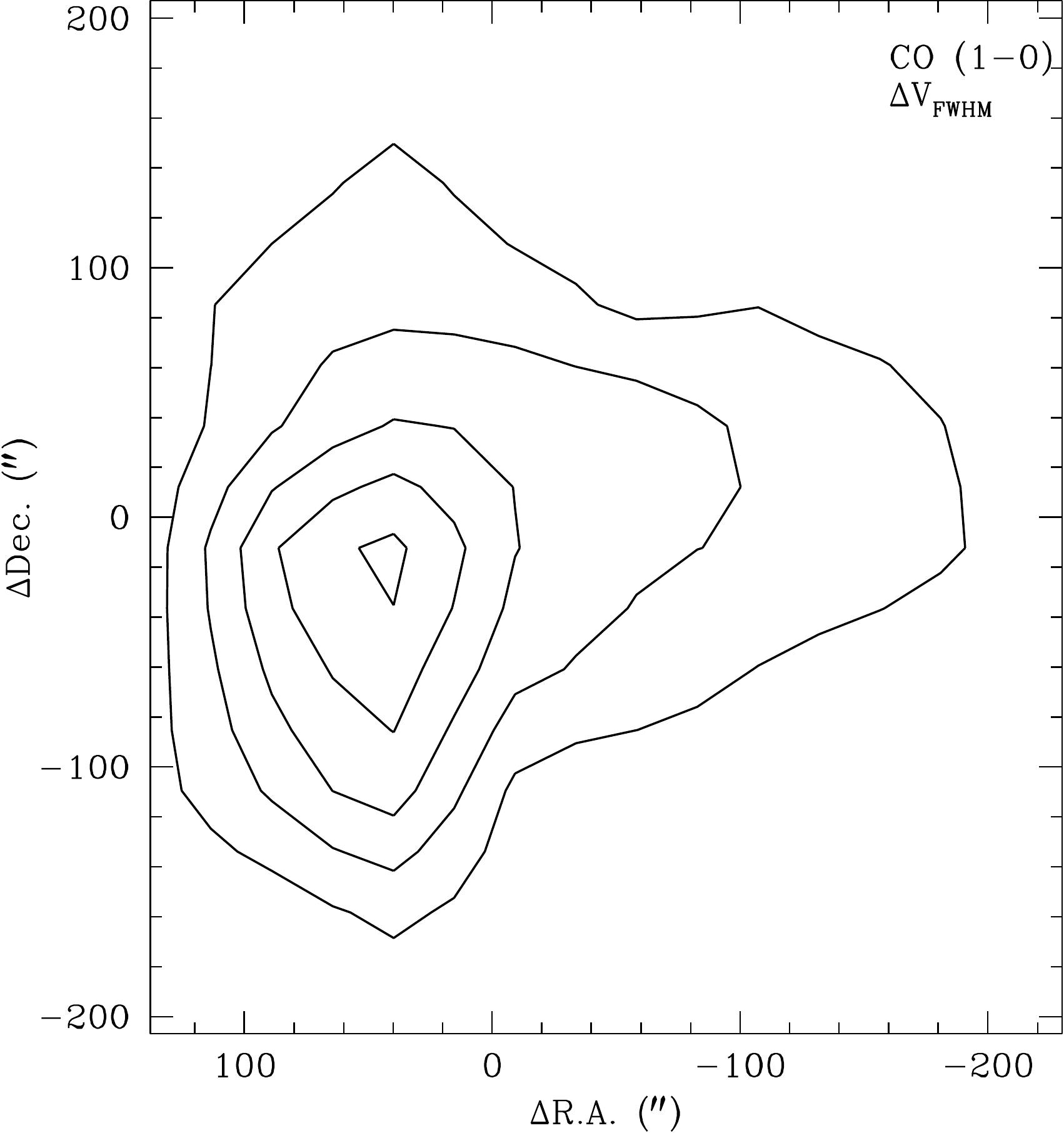}
\caption{
Smoothed contours of $^{12}$CO(1$-$0) line widths observed in the same region as Fig.~\ref{maps}.
The lowest contour is at 0.6 km~s$^{-1}$. Subsequent contours are in steps of 0.6 km~s$^{-1}$. 
The IRAS source is at the (0,0) position.
The PY10 dense core centre, at the $(+46,0)$ position, is also where the line width peaks.
}
\label{widths}
\end{figure}

Figure~\ref{widths} shows the contour map of the $^{12}$CO(1$-$0) line widths. The line widths across the cloud vary from 1.0~km~s$^{-1}$, at the positions where the gas is more quiescent with narrow lines well fitted by Gaussians, to 3.4~km~s$^{-1}$ at the PY10 core centre where a peak is located. At the (0,0) IRAS position, the line width is 1.8~km~s$^{-1}$, close or intermediate to the value of the quiescent gas. Thus, interestingly, the line widths at the PY10 core centre experience a large increase when compared with neighbouring positions, implying unusual dynamical activity, possibly with energetic embedded sources stirring up the gas.

Taken together, these results suggest that the star formation taking place in this small molecular cloud has lead to the formation of a group of young stars, the optical nebula stars, at the location of the IRAS source. As a result, the molecular gas at this position has been spent and dissipated. However, to the east there is much denser molecular gas, peaking at the PY10 core centre, about 46$''$ east of the IRAS source, and showing signs of dynamical activity and stirring, possibly caused by additional optical invisible sources, thus indicating current star formation activity, as demonstrated below.

\subsubsection{Molecular cloud mass and optical extinction} \label{cloud}

A few different methods can be applied to estimate the cloud mass using line 
emission of molecular tracers, in this case the CO isotopes.
A first estimate can be made using the empirical 
correlation between the $^{12}$CO(1$-$0) integrated intensity and
the H$_{2}$ column density along the line of sight, $N_{i,j}$, at position $(i,j)$:
\begin{equation}\label{nh2}
N_{i,j}(\mathrm{H}_2)=X \int T_{\mathrm{MB}}~dv \qquad (\mathrm{cm^{-2}})\;,
\end{equation} 
where the constant $X$ has been determined empirically. 
We use the value of $1.9 \times 10^{20}$~cm$^{-2}$~(K~km~s$^{-1}$)$^{-1}$
derived by \citet{strong96}.
To obtain the mass from the column density, we also used the ``typical'' value of 2.72 \citep{all73} for the mean molecular mass $\mu$. 
We obtain a mass of $M_{\mathrm CO}=1.6 \times 10^3 M_{\odot}$ for the mapped area.

Furthermore, the derived $N(\mathrm{H}_2)$ value of the column density towards any map position can be used to calculate the optical extinction $A_V$ \citep[see][]{bohlin78} towards that position. For the $(0,0)$=ESO~368-8/IRAS position and for the $(+46,0)$= PY10 core centre position, using the relation
$N(\mathrm{H}_2)/A_V=9.4 \times 10^{20}$~cm$^{-2}$~mag$^{-1}$ \citep{frerking82}, we obtain $A_V(0,0)=4.5$~mag and $A_V(+46,0)=15$~mag, respectively. 

Another method commonly used to estimate the cloud mass consists of assuming LTE conditions and using both an optically thick line and an optically thin line, in this case $^{12}$CO(1$-$0) and $^{13}$CO(1$-$0),
respectively. In this method \citep[see e.g.][for details]{pin08}, the peak main beam temperature of the $^{12}$CO(1$-$0) line is used to derive its excitation temperature, which, for example, is found to be 12.5~K at the map reference
position. Then, assuming that excitation temperatures are the same 
for both lines, the optical depth and the column density of $^{13}$CO
are calculated for each line of sight.
Finally, to calculate the total mass, it is necessary to adopt a H$_2$/$^{13}$CO
abundance ratios. The ``typical'' value is $5 \times 10^5$ quoted by \citet{dic78}.

The obtained mass amounts to  
$M_{^{13}\mathrm{CO}(1-0)}=6 \times 10^2$~$M_{\odot}$, which is compatible with what was obtained by means of the previous empirical method. 
The remaining discrepancy can be explained taking into account that the number of $^{13}$CO(1$-$0) spectra, contributing to the total map mass estimate, is markedly smaller than in the case of $^{12}$CO(1$-$0), due to smaller spatial coverage and lower $S/N$, and considering that the LTE approximation method seems to typically underestimate the $^{13}$CO true column densities \citep{pad00}. Finally, we should note that these two lines, $^{12}$CO(1$-$0) and $^{13}$CO(1$-$0), map different parts of the cloud because of their different optical depths, and also that, evidently, the empirical method based on the $^{12}$CO(1$-$0) emission is quite reliable in a statistical way, but for individual clouds it may give inaccurate results. Therefore, given these considerations, the two mass estimates derived here can be considered in good agreement.

The column density of $^{13}$CO can also be used to estimate the optical extinction $A_V$ towards the cloud by applying a linear relation: 
\begin{equation}
A_V=c_1~N(^{13}{\rm CO})+c_2.
\end{equation}

\noindent
Several estimates of the $c_1$ and $c_2$ 
parameters can be found in the literature, varying with the observed region 
and with the calibration technique. Here, we adopt the results of 
\citet{frerking82} for the Taurus region, $c_1=7.1 \times 10^{-16}$ mag~cm$^{-2}$ and $c_2=1.0$~mag.
The obtained visual extinction estimates are $A_V=3.6$~mag for the $(0,0)$ position and 12~mag for the $(+46,0)$ position, similar to the values obtained from the $^{12}$CO(1$-$0) data.



\subsection{The young stellar population}

\begin{figure*}[ht]
   \centering
   \includegraphics[width=\textwidth]{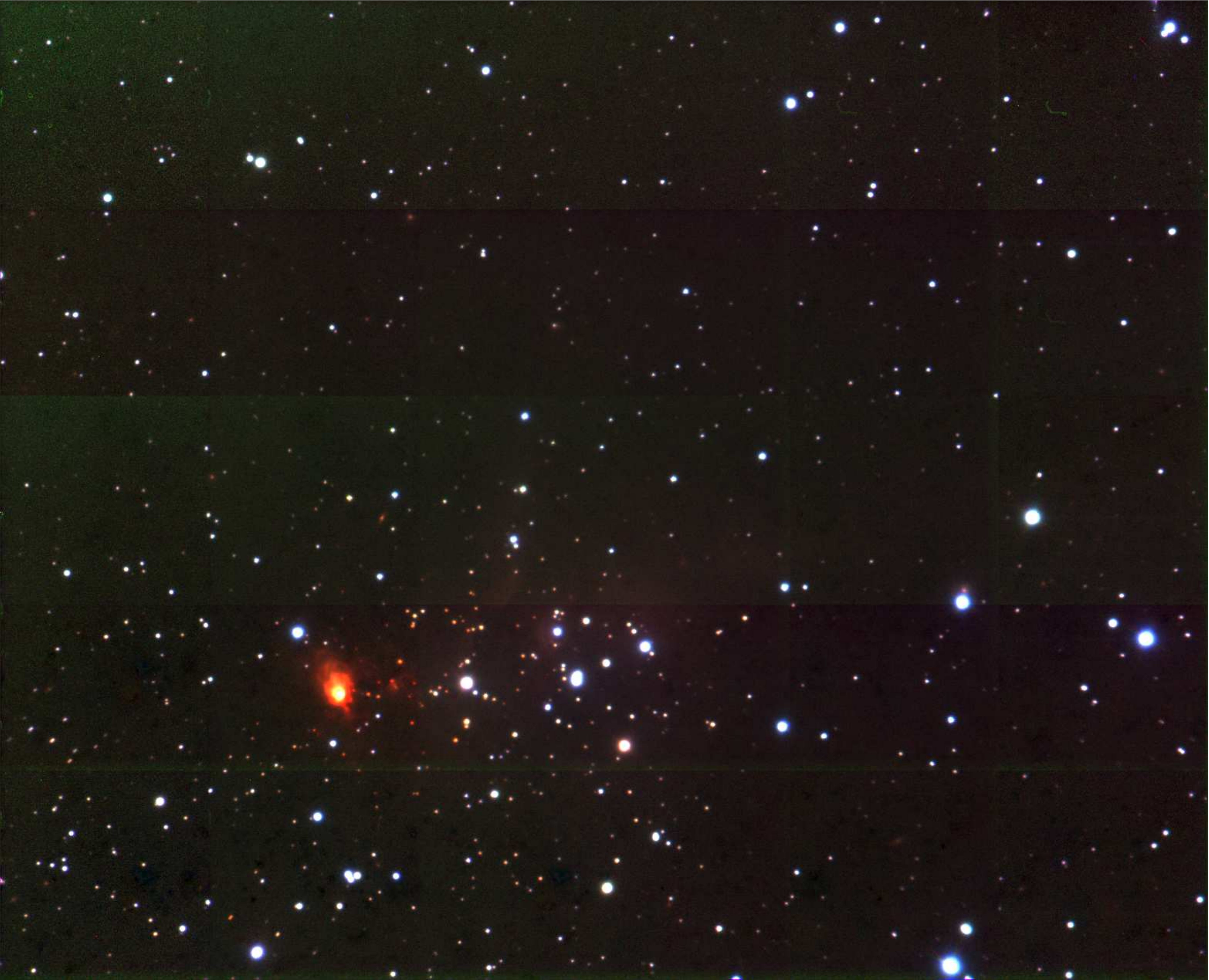}
   \caption{
$J$ (blue), $H$ (green), and $K_S$ (red) colour composite image towards ESO~368-8/IRAS~07383-3325 located in the region of the three optical nebula stars (south of centre). Notice the bright and very red near-infrared nebula (east of centre). The image covers about $3.8\times 3.1$ arcmin$^2$. North is up and east to the left.
	} 
	\label{JHK}%
    \end{figure*}

Figure~\ref{JHK} presents the ISAAC $JHK_S$ colour composite image obtained towards ESO~368-8/IRAS~07383-3325.
The three optical nebula stars are seen as relatively blue objects slightly below the centre of the image. However, the most striking aspect of this figure is the red nebula eastward of the centre of the image. This infrared nebula, very bright in the $K_S$-band and invisible at optical wavelengths, has the shape of a fan (or of a flame) and is analysed in more detail in the next section (\ref{IRnebula}).

\begin{figure}
   \centering
   \includegraphics[width=8cm]{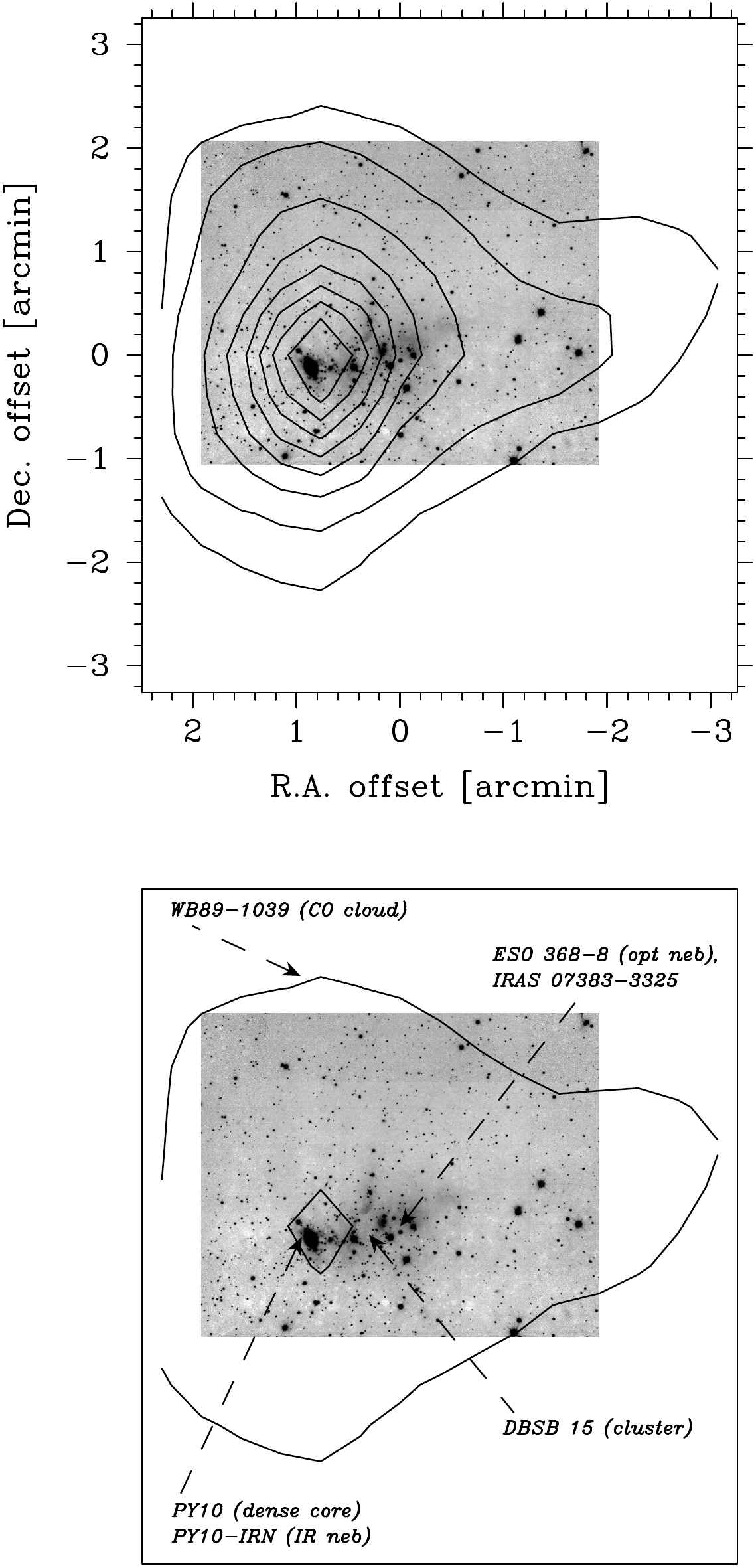}
   \caption{
{\it (top:)} Smoothed contours of $^{12}$CO(1$-$0) integrated intensity superimposed on the $K_S$-band image. The axes give the coordinates relative to the IRAS source. {\it (bottom:)} The spatial position of all the relevant objects: 1) the WB89-1039 CO cloud delineated by the lowest contour of our CO map, 2) the PY10 dense core delineated by the highest CO contour, 3) the DBSB~15 cluster, 4) the ESO 368-8 optical nebula coincident with the IRAS source, and 5) the new PY10-IRN near-infrared nebula.
	} 
	\label{co_K2}%
    \end{figure}

In Fig.~\ref{co_K2} (top), we show the relative positions of the molecular material (traced by the contour lines of the CO integrated intensity) and of the stellar content of this region as seen in our deep K-band image (greyscale). We notice the good spatial coincidence of the IR nebula and the PY10 dense core. The lower panel of Fig.~\ref{co_K2} displays the relative positions of all astronomical sources mentioned here.

%

Figures~\ref{JHK} and \ref{co_K2} also suggest a slight increase in the surface density of stars in the region surrounding the optical nebula stars and the infrared nebula, when compared with the rest of the image. In addition, there is clearly a large concentration of ``red'' stars in this region (much brighter in the $K_S$-band than in the $J$ or $H$-bands). These aspects will be quantified in section \ref{IRnebula} but, taken together, they suggest a possible small stellar cluster embedded in a region of local enhanced extinction.

%

\subsubsection{The infrared nebula} \label{IRnebula}

\begin{figure}
   \centering
   \includegraphics[width=7cm]{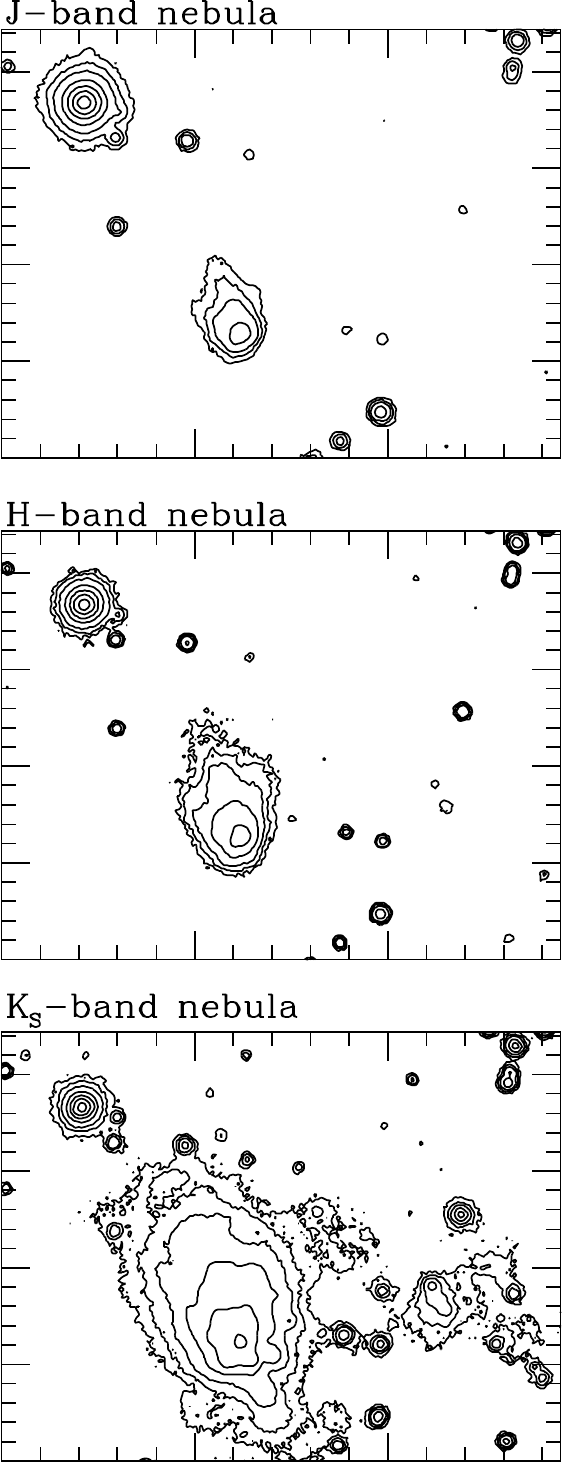}
   \caption{
Isophotes of the $JHK_S$-band images towards a newly discovered near-infrared nebula (``PY10-IRN'', coincident with the PY10 core centre, located about $1'$ east of ESO~368-8/IRAS~07383-3325). The small ticks on the axes indicate the size of 1$''$. North is up and east to the left. Contour levels are at 5, 10, 20, 60, 180, 600, and 2400$\sigma$ above the sky background.
	} 
	\label{neb}%
    \end{figure}

In Fig.~\ref{neb}, we present the isophotes of the $JHK_S$-band images, zoomed-in to show the details of the extended nebular emission seen about 46$''$ east of the IRAS source. This location coincides with the peak of the molecular emission and of the CO line width map 
(the cloud centre or PY10 core centre). Thus, we name this IR-nebula ``PY10-IRN''.

%

PY10-IRN is very bright and extended in the $K_S$-band image, becoming fainter and shrinking in size as we progress towards the $H$ and the $J$ bluer wavelength images. In each of the near-IR images, the nebula does not contain a detected point-source because the full width at half maximum (FWHM) at any point within the nebula is much greater than the FWHM of the real point sources measured across the images. This indicates that we are not seeing the PY10-IRN nebula illuminator source and that there is no direct line-of-sight to it.
Instead, the nebulosities seen in the images are composed of light scattered off the walls of cavities carved in the cloud by jets or outflows and into our line-of-sight. 

During the star formation process, the envelope and the molecular cloud core where a young embedded source resides is excavated by a stellar jet and molecular outflow, which are typically most intense during the earliest evolutionary stages of YSOs: Class~0 \citep{andre93}, and Class~I \citep{adams87}. In the Class~0 stage, the source still has to accrete the bulk of its final mass. In the Class~I stage, the source is accreting matter from a circumstellar optically thick disc and is also surrounded by an infalling envelope. 
The presence of these cavities allows radiation from the central embedded object to escape via scattering off the walls of the cavities. Depending on the inclination angle of the cavity axis of symmetry to the line-of-sight, different shapes result for the nebulae that have been found associated to young embedded sources. This can be seen in models of embedded infrared nebulae that have been produced by \citet{lazareff90}, \citet{whitney93}, \citet{whitney03}, and \citet{stark06}.
The shape of our IR-nebula can be compared with these models. It most resembles models no.\ 13 or 14 of \citet{whitney93} (their Fig.~4b), which corresponds to an embedded source in a cilindrical cavity with moderate optical depth and an inclination angle (of the polar axis) to the line-of-sight of $\cos i = 0.6$. Similarly, the \citet{stark06} models suggest that there is a curved or conical outflow cavity viewed at about 60$^{\circ}$ inclination.

A simple astrometric analysis of the location of the peak intensity of the nebula in each of the $JHK_S$ images supports the idea that we are seeing scattered light. The positions of the peaks in the $JHK_S$ images are not coincident. Instead they are offset by about 0.12$''$ and lie close to a straight line, with the values of the offsets correlating with the wavelength (i.e., the $H$-band peak is offset about 0.12$''$ from the $K_S$-band peak and the $J$-band peak is offset about 0.12$''$ from the $H$-band peak approximately in a straight line). The direction of this line is remarkably similar to the direction of the elongation of the $K_S$-band nebula. This behaviour is indicative of scattered light.

In order to provide an estimate of the relative brightnesses of the nebula in the $JHK_S$-bands, we performed aperture photometry of the nebula, for a circular aperture centred on the location of the peak emission, and an aperture radius equal to the one adopted for the true stars in the images. The results are:
$J_{\mathrm{(PY10-IRN})}=16.9$, $H_{\mathrm{(PY10-IRN})}=15.1$, and $K_{S\mathrm{(PY10-IRN})}=13.7$. Assuming that the coulours of the nebula are the same as the colours of its illuminator embedded YSO, we obtain $(J-H)=1.8$, $(H-K_S)=1.4$. These colours are also compatible with the source being in the Class~I stage of low-mass YSOs \citep{lada92}.

\subsubsection{Reddening and the embedded stellar population} \label{reddening}

\begin{figure}
   \centering
   \includegraphics[width=8cm]{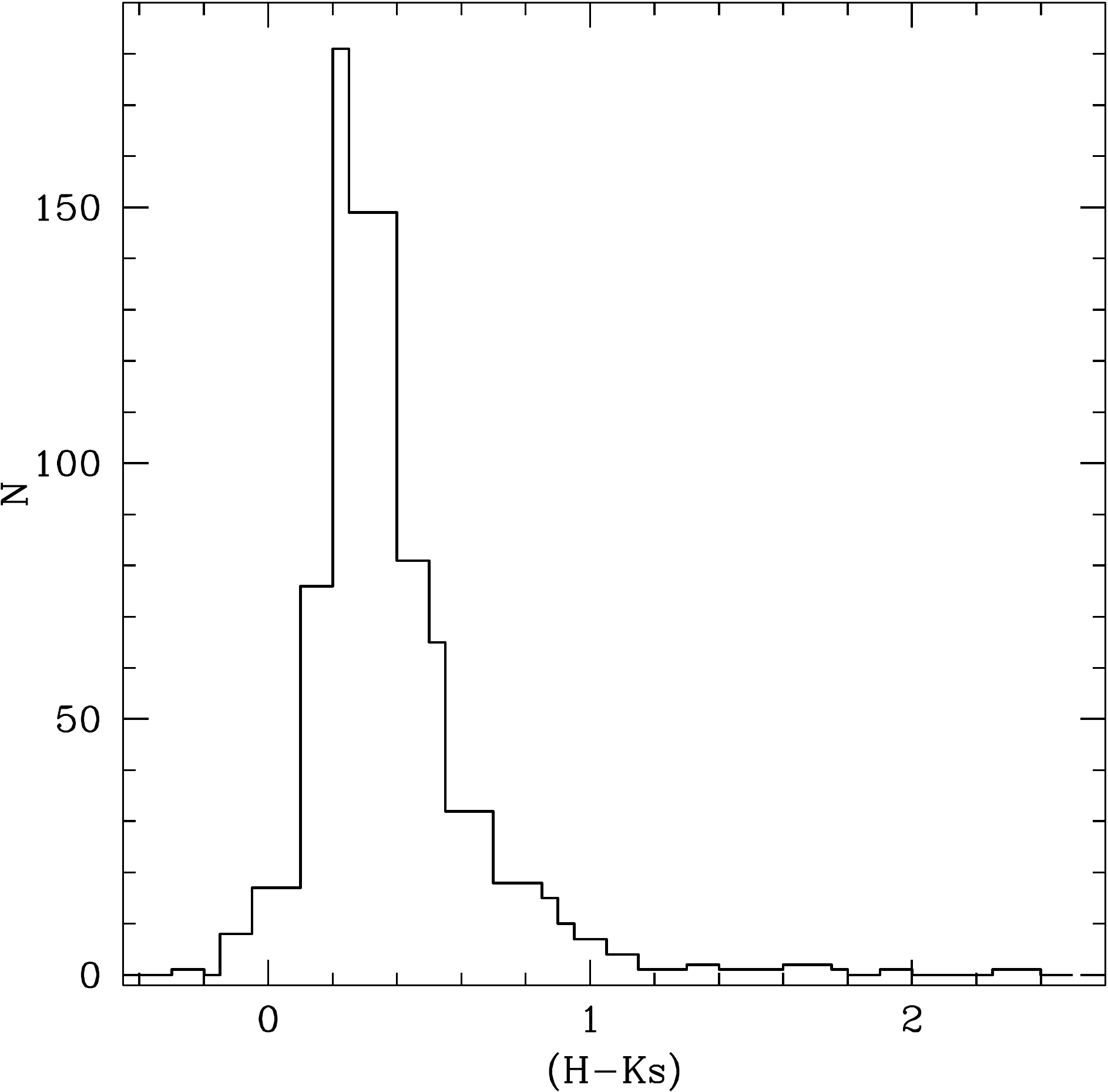}
   \caption{
Histogram of the observed $(H-K_S)$ colours. The well-defined peak is composed of foreground field sources (``blue'' sources with colours $(H-K_S)\le0.8$). The red wing of the distribution is composed of sources spatially concentrated in the region where the molecular cloud is present (see Fig.~\ref{scatter}). 
	} 
	\label{histHK}%
    \end{figure}

\begin{figure}
   \centering
   \includegraphics[width=8cm]{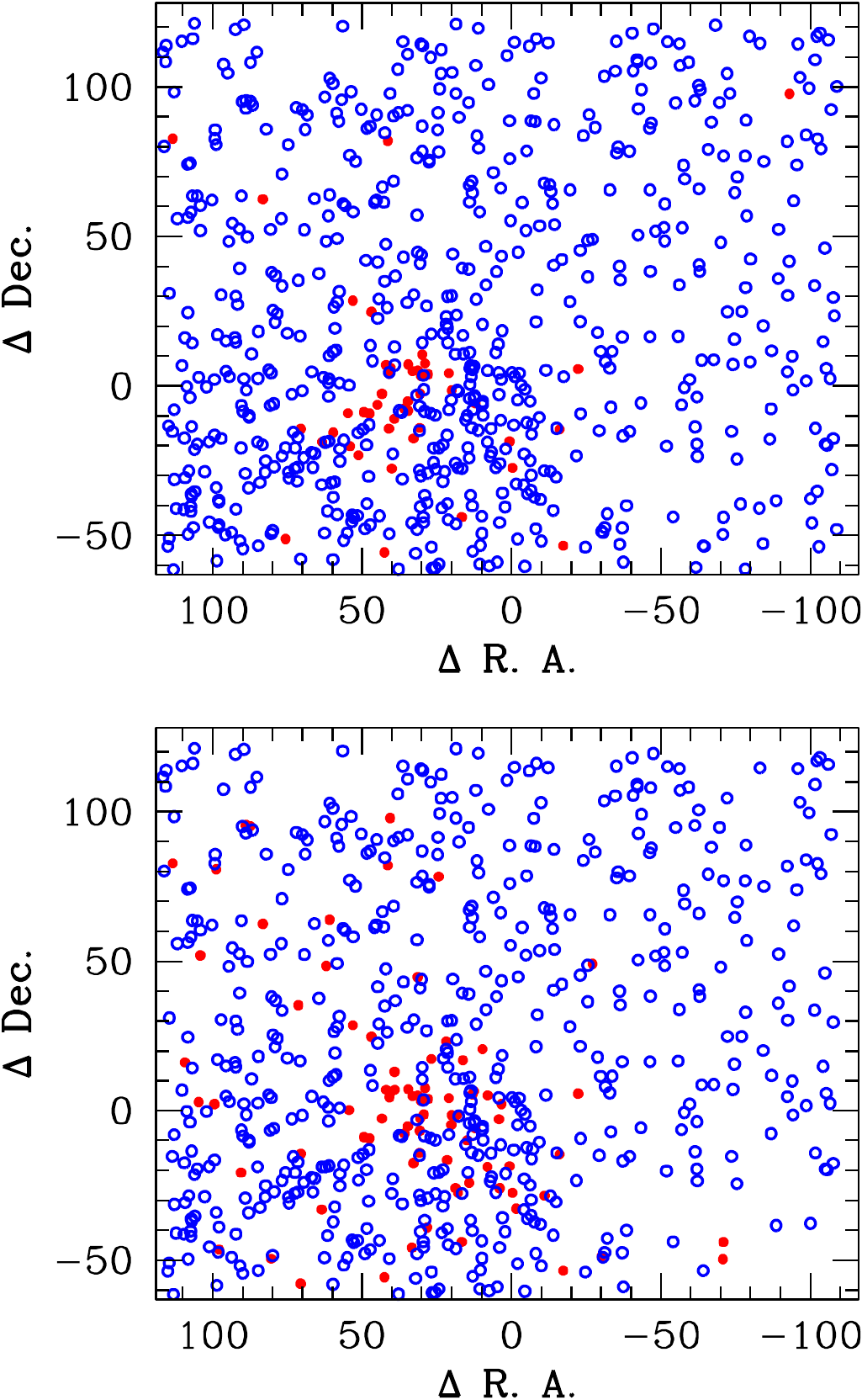}
   \caption{
{\it (top:)} Spatial distribution of all the sources seen both in the $H$ and in the $K_S$ band images. Blue open circles represent sources with values of  $(H-K)$ lower than 0.8.  Red filled circles represent sources with values of $(H-K)$ higher than 0.8.  This plot is centred on the $IRAS$ point source. 
{\it (bottom:)} Spatial distribution of all the sources seen simultaneously in the $J$, $H$ and $K_S$ band images. Red filled circles represent sources to the right of the reddening band (see next figure). Blue open circles represent sources located inside the reddening band.
	} 
	\label{scatter}%
    \end{figure}

Figure~\ref{histHK} shows the histogram for the observed $(H-K_S)$ colours of the sources detected in both the $H$ and the $K_S$-band images. In the range of 0 to about 0.8, the curve is close to symmetric and most sources here  correspond to field main-sequence stars (located in front of the molecular cloud). The spread in $(H-K_S)$ values, about the peak value of $(H-K_S)=0.35$, stems from the range of intrinsic colours of main-sequence stars and to low values of variable foreground extinction in the lines of sight of each source. The mean value of the $(H-K_S)$ colours within this peak (for $(H-K_S)\le0.8$) is 0.33, with a standard deviation of 0.15. 
The spatial location of these peak ``blue'' sources is seen in Fig.~\ref{scatter} (upper panel) where they are represented by blue open circles. Red filled circles, on the other hand, represent sources in the red tail of the $(H-K_S)$ histogram, the ``red'' sources. The blue sources are scattered randomly and uniformly across the image, whereas the red sources are concentrated in the region of the molecular cloud. Taken together, these results strongly indicate that most red sources are objects associated to the cloud, either located behind the cloud or being embedded in the cloud and possibly containing near-infrared excess emission from circumstellar material. 
We argue here that, due to the location of this molecular cloud well below the Galactic plane and the sparsity of stars in the Galactic halo, few sources seen in the near-infrared images are background sources. 

%

For the general case, this colour excess of $(H-K_S)=0.8$ (about equal to the mean value of $(H-K_S)$ for the field sources plus 3 standard deviations)  would just separate the objects into two groups: foreground stars and \{embedded + background\} stars. However, given that there are very few background stars at this location below the Galactic plane,
this reddening effect
separates foreground from embedded stars and thus effectively selects YSOs embedded in the cloud.
Thus, the large majority of the red sources are likely to be YSOs embedded in the molecular cloud.

%

Blue sources, on the other hand, may be composed of a mix of foreground field sources and YSOs in a more evolved evolutionary stage. These more evolved young stars, if present, could be pre-main-sequence objects or even intermediate-mass or massive main-sequence stars formed in this cloud, which evolve much faster than their lower-mass siblings formed at the same time. Their higher masses would also contribute to their being bluer and thus not being told apart by red colours.

\subsubsection{Young stellar objects}

   \begin{figure}
   \centering
   \includegraphics[width=8cm]{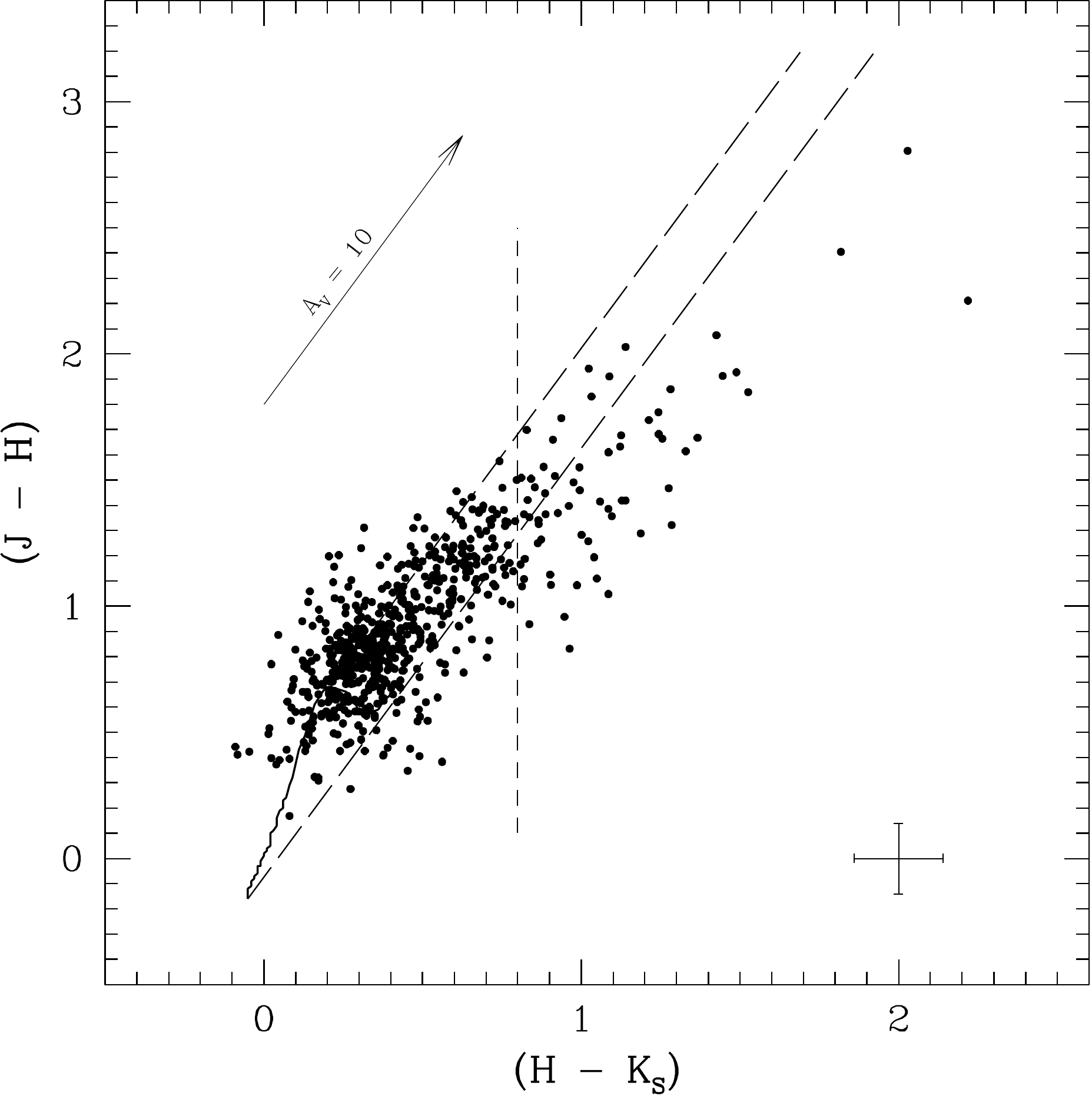}
   \caption{
Near-infrared colour-colour diagram of the region towards IRAS~07383-3325. 
The solid line represents the loci of unreddened main-sequence stars \citep{bessel88}, while long-dashed lines indicate the reddening band. The reddening vector indicates the direction of the shift produced by extinction by dust with standard properties.
The location of the vertical dashed line, derived from Fig.\ref{histHK}, is at $(H-K_S) = 0.8$. The cross in the lower right corner represents a typical error bar.
	} 
	\label{cc}%
    \end{figure}
%

Using the 648 point sources detected in all three $J$, $H$, and $K_S$-bands, 
we plotted the near-infrared colour-colour diagram, $(J-H)$ versus $(H-K_S)$, shown in Fig.\ref{cc}.
Most stars are located within the reddening band where stars appear if they are main-sequence stars reddened according to the interstellar extinction law \citep{rieke85}, which defines the reddening vector (traced here for $A_V=10$).
Pre-main-sequence YSOs, or massive main-sequence stars recently formed in this region, which have had time to clear the inner regions of their circumstellar discs, lie in this region as well. Giant stars appear slightly above this band. Stars located to the right of the reddening band are likely to be embedded young star objects with infrared excess emission from circumstellar material \citep{adams87}. 

For the sources that lie inside the reddening band, the highest value of $(H-K_S)$ is about 1.14. Using the mean value of $(H-K_S)=0.33$ for field stars (according to Fig.~\ref{histHK}), we obtain a colour excess $E(H-K_S) = 0.81$ due to intra-cloud extinction, which corresponds to about a maximum visual extinction produced by the cloud core of about $A_V=13$ \citep{rieke85}. This value is very similar to those derived towards the densest part of the cloud (position (+46$''$,0) or PY10 core) using CO spectra (see section \ref{cloud}).

The location of the vertical dashed line, derived from Fig.~\ref{histHK}, is at $(H-K_S) = 0.80$. The two groups of sources, blue sources with $(H-K_S) < 0.80$ and red sources with $(H-K_S) \ge 0.80$, are very differently distributed on the colour-colour diagram.  A large fraction of the red sources are located outside and to the right of the reddening band, whereas the blue sources mostly occupy the inside of the reddening band. Thus, most red sources are likely to be YSOs. Given their spatial concentration (Fig.~\ref{scatter}), these red sources, with $(H-K_S) \ge 0.80$, seem to represent a small young embedded stellar cluster 
of about 46 young stars forming in the molecular cloud. 
The actual number of stars formed in this cloud is likely to be larger for at least three reasons. Firstly, we chose a conservative value of $(H-K_S) = 0.80$ ($3\sigma$ above the mean value of the $(H-K_S)$ of field main-sequence stars). Secondly, the presence of the optical nebula stars seems to indicate that there are some young stars in a more advanced stage of the star formation process, already free of circumstellar material and exhibiting blue colours, thus not pinpointed by our colour selection criterion. Thirdly, there are IR-excess sources with $(H-K_S) < 0.80$ that are YSOs belonging to the cluster.
The presence of this YSO population is likewise supported by Fig.~\ref{scatter} (lower panel) where a plot similar to the upper panel is shown. We notice that unlike the upper panel, in the lower panel red filled circles represent sources with IR excess emission (derived from Fig.~\ref{cc}). The distribution of the IR excess stars is similar to the distribution of the large $(H-K_S)$ sources. Both are concentrated at the location of DBSB 15, between ESO368-8/IRAS~07383-3325 and the PY10-IRN nebula (closer to the latter), and trace the young stellar population forming in this cloud core. Thus, as mentioned above, the specific location of this SFR in the Galaxy makes the two approaches illustrated in Figs.~\ref{scatter} almost equivalent in pinpointing young stars.

Given the uncertainty in the number of stars forming in the cloud, we can make a rough estimate of the star formation efficiency of this molecular cloud. Assuming that there are 46 solar mass stars and 3 optical nebula stars that could reach about 6 solar masses each (see next section), we obtain a total stellar mass of $\sim 64 M_{\odot}$. For a total mass of the gas in the cloud between 600 and 1600~$M_{\odot}$, we obtain a star formation efficiency in the range of 4--10\%. Even though the uncertainty is large, these values are similar
to the lower end values of star formation efficiencies found in cluster environments within the local star formation regions \citep[e.g., L1630;][]{lada99}.

\section{Discussion}

The two non-spatially coincident nebulae, ESO~368-8 (seen in the optical) and PY10-IRN (seen in the near-IR images), together with a small enhancement of the surface stellar density are the main signs of star formation in the form of a small embedded young cluster in the WB89-1039 molecular cloud. Several aspects make this region particularly interesting. 
The first aspect is the Galactic location, in the outer Galaxy and well below the Galactic plane. Even though not in the far outer Galaxy \citep[defined as Galactocentric distances $>$13.5 kpc; this number is based on the radial distribution of CO emission from the studies of][]{digel96,heyer98}, this location shares at least one aspect with the far outer Galaxy: it is close to the ``edge'' of the Galactic molecular disc (in this case, the lower edge), about 500~pc (vertical distance) below the plane of the Galaxy defined by $b=0^{\circ}$.
Interestingly, star formation occurs frequently and with no major differences in these edge environments, both in the far outer Galaxy \citep[e.g.][]{brand07,yasui08,yun09}, and also well below the Galactic plane as shown here. The relatively small number of molecular clouds at this large vertical distance \citep{may97} do not appear to significantly affect the occurrence and properties of the star formation process. The only significant difference of these ``edge'' star formation sites is that they do not seem to form massive stars \citep[e.g.][]{yun07,yun09}. In fact, based on the observed fluxes, no massive star is expected to be forming in this region.

In the following, we discuss the nature of the optical nebula stars and the presence of the two non-spatially coincident nebulae, focusing especially on what they suggest for the star formation scenario in this molecular cloud.

\subsection{The optical nebula stars}

\begin{figure}
   \centering
   \includegraphics[width=8cm]{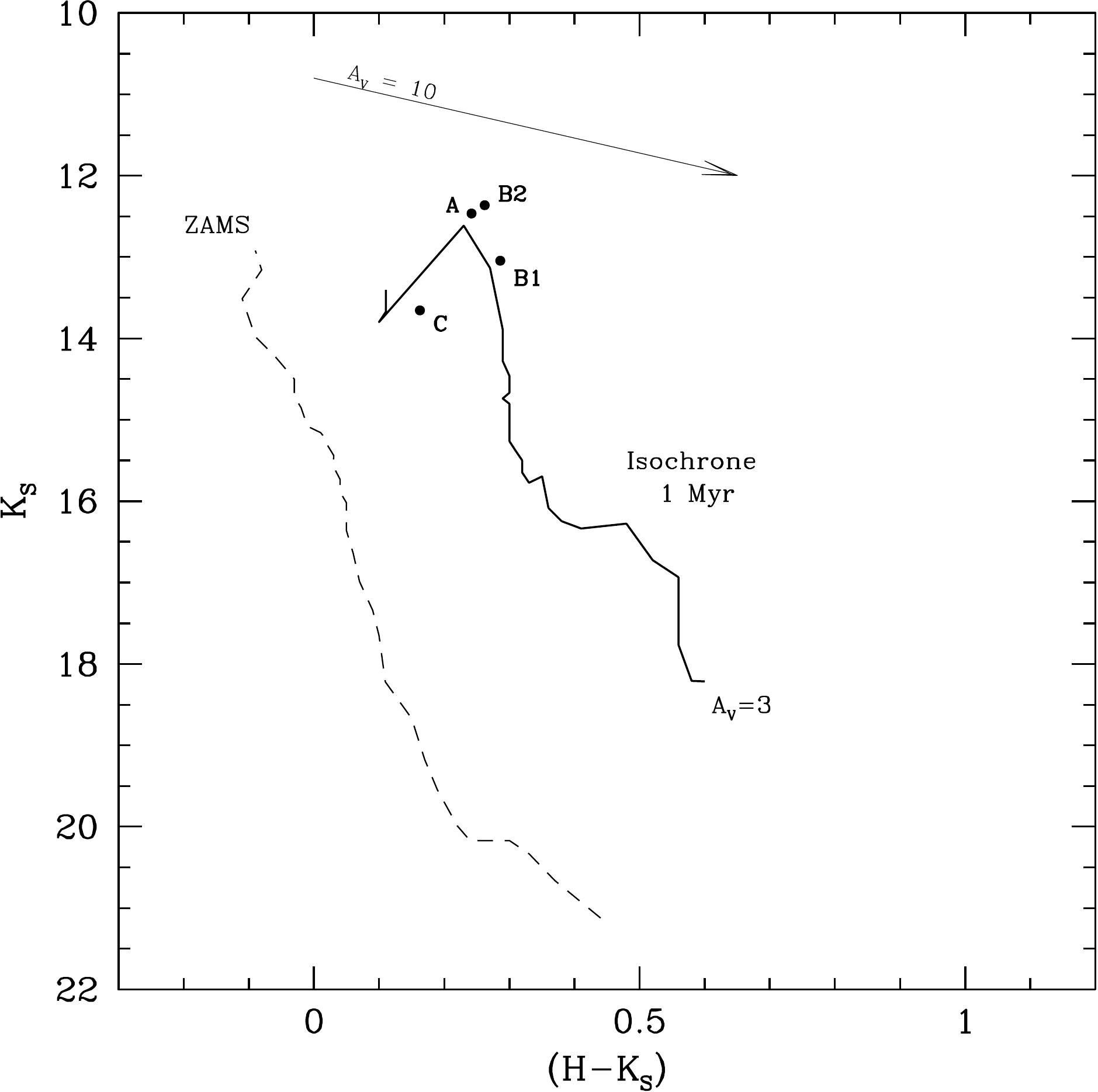}
   \caption{
Colour-magnitude diagram with the positions of the optical nebula stars (A, B1, B2, and C) marked by filled dots. The dashed line is the zero-age main-sequence \citep[ZAMS;][]{siess00} for the cloud distance and with zero extinction. The solid line is the 1 Myr isochrone with extinction $A_V=3$. The reddening vector indicates the direction of the shift produced by extinction by dust with standard properties \citep{rieke85}.
	} 
	\label{cm}%
    \end{figure}

The nature of the stars involved in the optical nebula can be investigated using the $(H-K_S),K_S$ colour-magnitude diagram seen in Fig.~\ref{cm}. In this figure, we have marked the positions of the optical nebula stars (A, B, and C). The optical nebula star B is in fact a double star resolved in our near-IR images (whose stars are named here B1 and B2). The best fit to these sources is achieved using the 1 Myr isochrone of \citet{siess00} with extinction $A_V=3$. This value is in good agreement with the value of $A_V$ derived towards these stars (the (0,0) position) using the CO data (Sect.~\ref{cloud}). Bearing in mind the relatively large uncertainties when fitting isochrones to colour-magnitude diagrams, we estimate that the optical nebula stars have an age close to 1 Myr and are intermediate-mass stars (from spectral type B5 to A9). Furthermore, using $A_V=3$, the absolute magnitudes in the $V$-band of the optical nebula stars span values that are a few magnitudes greater than the corresponding values for intermediate-mass main-sequence stars. This is compatible with these stars being pre-main-sequence intermediate-mass stars whose luminosities are still higher than the corresponding ones when they reach the main-sequence.

%

\begin{figure}
   \centering
   \includegraphics[width=8cm]{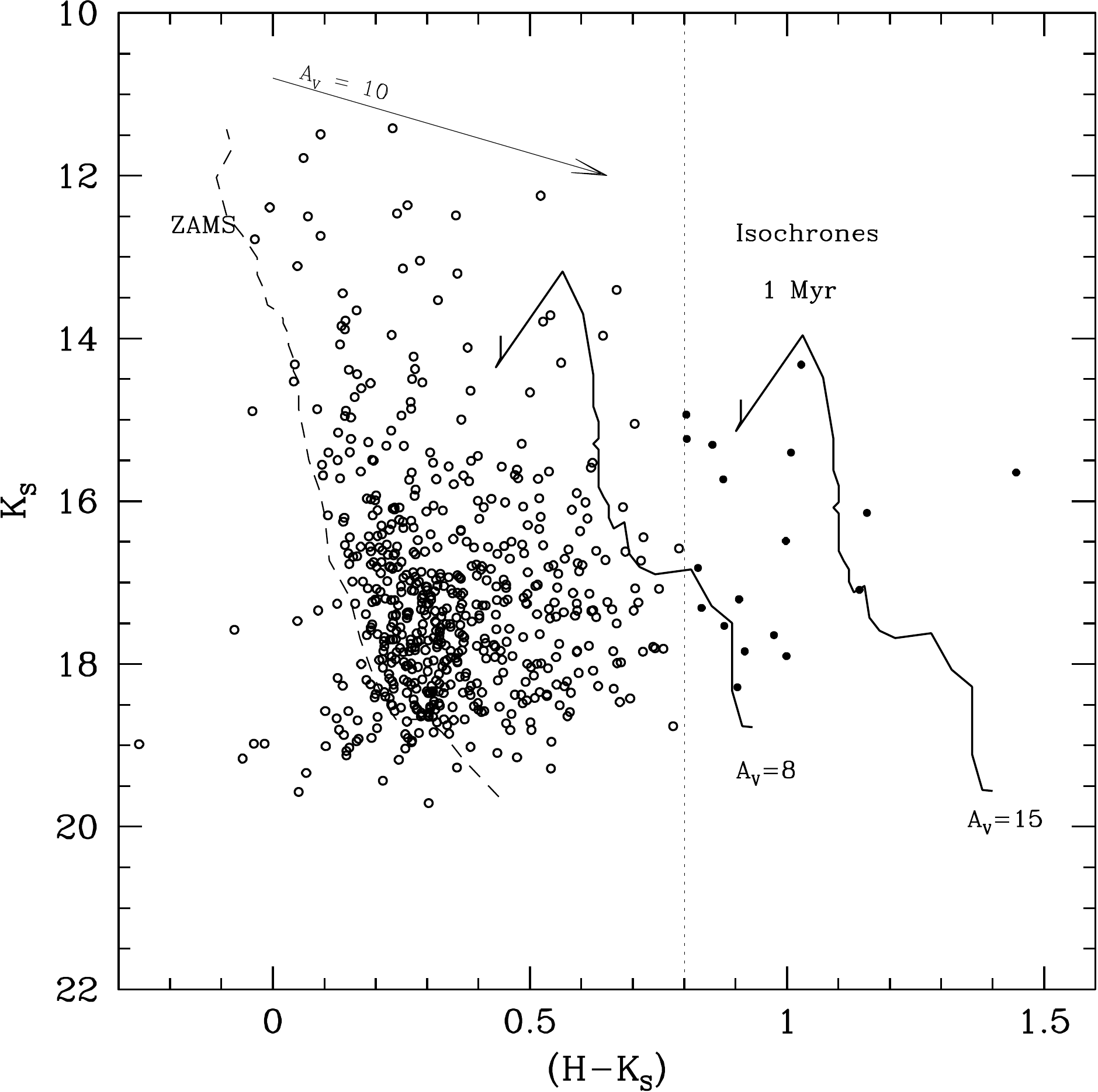}
   \caption{
Colour-magnitude diagram for the sources seen in both the $H$ and the $K_S$-bands with no signs of excess emission due to circumstellar material. The dashed line is the zero-age main-sequence for an arbitrary distance. The two solid lines are the 1~Myr isochrones with extinctions $A_V=8$ and $A_V=15$, respectively. Filled circles represent the red sources with $(H-K_S) \ge 0.8$. Open circles represent the blue sources with $(H-K_S) < 0.8$. 
	} 
	\label{isoHK2}%
    \end{figure}

Figure~\ref{isoHK2} shows the colour-magnitude diagram for the sources seen in both the $H$ and the $K_S$-bands but excluding those that exhibit signs of excess emission due to circumstellar material (as inferred from Fig.~\ref{cc}).
In this way, the location of a star on this diagram depends on its distance (vertical shift), on its extinction (shift along the reddening vector) and on its age (shift given by pre-main-sequence theoretical models of stellar photospheres with no circumstellar emission).
The two groups of sources (filled red sources with $(H-K_S) \ge 0.8$ and open blue sources with $(H-K_S) < 0.8$) are separated by the vertical dotted line. The blue sources represent mostly foreground main-sequence stars, at variable distances and small extinction. The red sources can be best fitted by a 1~Myr isochrone with variable values of extinction, as is known to exist toward embedded sources, between about $A_V=8$ and $A_V=15$, consistent with the values of $A_V$ towards these sources (mostly located around the $(+46,0)$ position) derived from the molecular data (Sect.~\ref{cloud}). Thus, this plot supports the idea that we are in the presence of a young stellar cluster or aggregate with about 1~Myr age and composed of at least some tens of stars.

%

\subsection{The two reflection nebulae}

The two reflection nebulae, the optical nebula ESO~368-8 and the near-IR nebula PY10-IRN, are separated by about 46$''$ (1.1 pc). Not surprisingly the location of the optical nebula is relatively poor in gas, when compared with the location of the near-IR nebula where the amount of gas (and the extinction) is much greater. Do the two nebulae mark two star formation sites in the same cloud, or can the two nebulae be considered coeval and two manifestations of the same episode of star formation taking place in the cloud? In the latter case, the two different stages of the young stars, the optical nebula stars and the red sources close to the near-IR nebula, could stem from different stellar masses. The optical nebula stars should be sufficiently more massive than the stars forming close to the near-IR nebula. Because of their higher masses, they would have evolved faster and dissipated the cloud material while their lower mass siblings to the east are still heavily embedded and extincted at optical wavelengths.

On the other hand, if the two nebulae represent two non-coeval star formation sites, could this be the case for triggered sequential star formation? 
Using the gas temperatures from the CO observations to derive the sound speed in the cloud, we conclude that it would take about 2.3 Myr for a pressure wave to travel from the optical nebula to the near-IR nebula (for a projected distance of 1.1 pc between the two nebulae). If the optical nebula stars have triggered star formation at the near-IR location, they had to have done it at least 2.3 Myr ago. This value is of the order or greater than the age derived for the optical nebula stars. This fact, and the ages derived in the previous section,  favour the scenario of the nebulae being considered coeval and representing two manifestations of the same single star formation episode.
Moreover, the higher velocities inferred from the line widths within the PY10 core, peak at the position of the near-IR nebula. This indicates that the source of gas stirring here is likely to be an embedded source at this position, not a source located far out at the position of the optical nebula where line widths are not enhanced.

\section{Summary}
 
We present a study of the molecular gas and young stellar population seen towards the region of the reflection nebula ESO~368-8 and the IRAS source IRAS~07383-3325, in the molecular cloud WB89-1039 located in the outer Galaxy and well below the Galactic plane.
We confirm the presence of a young stellar cluster (or aggregate of tens of YSOs), identified by \citet{dutra03}, composed of low and intermediate-mass stars, with no evidence of high-mass stars.

This star formation region is associated with the ESO~368-8 optical reflection nebula and with a near-infrared nebula (PY10-IRN) located about 46$''$ (1.1~pc) from each other. The two nebulae seem to be coeval and to represent two manifestations of the same single star formation episode with about 1~Myr age.
The near-IR nebula appears to be composed of scattered light off a cavity carved by previous stellar jets or molecular outflows from an embedded, optically, and near-IR invisible source.

The molecular cloud was fully covered by our CO($J$=1$-$0) maps and, traced by this line, extends over a region of $\sim 7.8 \times 7.8$~pc$^2$, exhibiting an angular size $\sim 5.4' \times 5.4'$ and shape (close to circular) similar to spherical (or slightly cometary) globules. 
Towards the direction of the near-IR nebula, the molecular cloud contains a dense core (PY10) where the molecular gas exhibits the large line widths indicative of a very dynamical state, with stirred gas and supersonic motions that could generate turbulence.

Our estimates of the mass of the molecular gas in this region range from 600 to 1600 solar masses.
In addition, the extinction $A_V$ towards the positions of ESO~368-8 (optical reflection nebula) and of PY10-IRN (near-IR nebula), derived by means of different column density methods, was found to be $A_V \simeq 3-4$~mag and  $A_V \simeq 12-15$~mag, respectively.
We estimate a value of $\sim 4-10$\% for the star formation efficiency of this molecular region.

\begin{acknowledgements}
This work has been partly supported by the Portuguese Funda\c{c}\~ao
para a Ci\^encia e Tecnologia (FCT) and by the European Commission FP6 Marie Curie Research Training Network ``CONSTELLATION'' (MRTN-CT-2006-035890).
The research made use of the NASA/IPAC Infrared Science Archive, which is operated by the Jet Propulsion Laboratory, California Institute of Technology, under contract with the National Aeronautics and Space Administration.
This research made use of the SIMBAD database, operated at CDS, Strasbourg, France, as well as SAOImage DS9, developed by the Smithsonian Astrophysical Observatory.
\end{acknowledgements}

\end{document}